# Three-axis Active Magnetic Compensation with Spatial Feedforward Matrix Decoupling and Hysteresis PI Regulation for Suppressing Zeeman Decoherence of Cold Atom Quantum Memories*

Ya Li(李雅)[1,2,3], Xiao-Long Zou(邹晓龙)[1], Xiang-Yao Liang(梁湘瑶)[1], Chang-Qing Niu(牛常清)[1], Xin Guo(郭鑫) [1,2], Wei-Zhe Qiao(乔伟哲)[4] and You-Xing Chen（陈友兴）[1,3**]

[1]School of Information and Communication Engineering, North University of China, Taiyuan 030051, China,

[2]Shanxi Province Key Laboratory of Intelligent Detection Technology and Equipment, North University of China, Shanxi Taiyuan 030051, China,

[3]State key Laboratory of Extreme Environment Optoelectronic Dynamic Measurement Technology and Instrument, North University of China, Shanxi Taiyuan 030051, China,

[4]Shanxi Dazhong electronic information Industry Group Co., LTD., Taiyuan 030024, China

[*]Supported by National Natural Science Foundation of China (Grant Nos. U23A20636, 62204232, 62601745), Fundamental Research Program of Shanxi Province (Grant No. 202303021222109), Natural Science Foundation of Shanxi Province (Grant Nos. 20210302124189, 202303021212208, 202403021222162), and Shanxi Scholarship Council Research Project (Grant No. 20210038).

[**]Corresponding author. E-mail: chenyouxing@nuc.edu.cn (You-Xing Chen, main corresponding author); liya@nuc.edu.cn (Ya Li)

---

Cold-atom quantum memories require a near-zero magnetic environment (residual field <1 μT) to mitigate Zeeman decoherence. Conventional three-axis active compensation, however, is limited by inter-coil crosstalk and geomagnetic drift, making it difficult to guarantee the uniformity and long-term stability of the zero-field region. To address this issue, we propose a spatiotemporally cooperative feedback-controlled suppression architecture for magnetic-field interference: an integrally printed three-dimensional orthogonal coil frame is adopted to mechanically suppress inter-axis flux leakage, a spatial feedforward inverse matrix is introduced to algorithmically eliminate first-order cross-coupling, and a dual-threshold hysteresis incremental-PI controller is designed to filter magnetometer quantization fluctuations and avoid frequent small jitter adjustments of the coil driving currents. Hardware ablation experiments, long-duration stability tests, and spatial magnetic-field mapping characterization were conducted. Experimental results show that during a selected 714.4 s continuous steady-hold stage, the mean and maximum residual fields at the central point reach 0.112 μT and 0.224 μT, respectively, and the overlapping Allan deviations of all three components remain below 5 nT for averaging times of 13.9–178.2 s. Based on drift-referenced 27-site lattice-scan measurements, combined with centre-constrained quadratic vector reconstruction and soft Maxwell regularization, a weak-field spatial distribution with an RMS variation of 5.65 μT can be fitted within a spherical region of interest of radius 5 mm, where 67.5% of the model volume satisfies the sub-5 μT coverage criterion; for a prolate ellipsoidal region corresponding to the atomic-cloud distribution with full axis lengths of 10, 5, and 5 mm, the upper field bound within 93.7% of the volume is 2.68 μT. The spatial distribution of the weak-field region, together with real-time measurement feedback, facilitates timely sensing and suppression of magnetic drift in the atomic region, thereby improving the temporal stability of the central field and the spatial uniformity within a finite volume, sustaining the coherence of stored spin waves over long times and enhancing the fidelity of subsequent entanglement. This architecture provides an engineering basis for DLCZ cold-atom quantum memories and can serve as a design reference for SERF magnetometers and space cold-atom clocks.

1. Introduction.

Collective atomic spin waves store quantum information in quantum repeaters, whose coherence lifetime determines the maximum entanglement transmission distance.[1,2] Ambient magnetic perturbations trigger Zeeman splitting and time-varying phase drift, causing the spin coherence to severely degrade[3]. For 87Rb, the linear Zeeman shift is approximately 7 kHz/μT, and a geomagnetic field of 50 μT alone can produce 350 kHz of splitting and induce rapid dephasing.[4,5] Although clock states can suppress first-order Zeeman noise, quadratic shifts and decoherence caused by magnetic gradients and slow magnetic fluctuations (SMFFs) remain unsolved.[6-9] Existing experiments indicate that gradient mismatch still markedly reduces entanglement storage efficiency even under optical-lattice confinement.[10-14] Sun et al. further confirmed that asynchronous power-supply-induced magnetic drifts cause inter-node phase offsets in DLCZ networks, thereby destroying single-photon entanglement.[9] Natural geological structures and laboratory electronic equipment likewise generate random background magnetic fluctuations.[15-17] Near-zero-field systems (residual field <1 μT) must therefore simultaneously provide a uniform weak-field region and real-time dynamic geomagnetic compensation capability.

Two mainstream schemes are adopted for near-zero magnetic environments: passive magnetic shielding and triaxial active Helmholtz compensation.[18-20] Passive metal shielding suffers high cost, blocked optical paths and incomplete residual noise suppression; uneven internal fields make it unable to maintain long-term stable ultraweak magnetic zones alone.[21-23] Triaxial active coils create open-space compensation fields compatible with cold atom optical layouts and have been applied in SERF magnetometers and ion traps.[24-26] Nevertheless, linear feedback amplifies digital gaussmeter quantization noise near zero field and generates high-frequency current ripples that disturb spin precession.[20,27] Existing SERF-oriented three-axis compensation and RC-controlled exponential coil signals only offer partial magnetic regulation solutions.[28,29]

Current optimization efforts usually upgrade either the coil hardware or the control algorithm separately.[30] Differential biplanar coils, symmetric/mirror/swarm-intelligence coil designs and sensitivity matrix calibration relieve inter-axis crosstalk and field inhomogeneity [31-35], while BP neural networks can fit nonlinear magnetic coupling.[36] However, conventional separated nested triaxial coils bring large orthogonal errors and narrow uniform volumes, making it difficult to satisfy the miniaturization requirements of space-borne cold-atom clocks and integrated SERF devices.[24,37,38]

To eliminate the crosstalk introduced by coil orthogonality and ambient magnetic disturbances, this work proposes a collaborative triaxial active compensation scheme that integrates square Helmholtz frames, a feedforward decoupling matrix, and independent incremental PI controllers. Mechanically, the monolithic skeleton structurally constrains orthogonal deviation and cross flux leakage. Algorithmically, least-squares multi-point calibration constructs the coupling matrix, and feedforward pre-compensation suppresses inter-axis interference, while three incremental PI loops dynamically counteract geomagnetic baselines and SMFFs. To this end, a closed-loop test platform integrating a three-axis gaussmeter and a multi-channel precision DC power supply was established for full performance characterization. Repeatability was assessed through three long-duration runs, and a near-matched matrix-ablation experiment examined the contribution of coordinate decoupling. Within the common 90–120 min time window, the time-weighted mean residual-field magnitude across runs was (0.879 ± 0.232) μT. A separate fixed-command scan sampled 27 positions on a three-dimensional lattice to evaluate finite-volume spatial variation. After drift-referencing, the measured vectors were reconstructed using a centre-constrained quadratic vector model with soft Maxwell regularization. Within the spherical region of interest of radius 5 mm, the reconstructed centre-referenced RMS variation is 5.65 μT, with 67.5% of the model volume satisfying the 5 μT spatial-variation criterion. For the elongated prolate ellipsoidal region of interest with full axis lengths of 10, 5, and 5 mm, the corresponding RMS variation decreases to 2.68 μT and the qualifying volume fraction reaches 93.7%. These spatial statistics depend on the reconstruction model and the selected region of interest, whereas the sub-1 μT requirement mentioned above refers to the temporal stability of the residual field at the central feedback position. The proposed spatiotemporal framework therefore simultaneously improves both the temporal stability of the central field and the finite-volume spatial inhomogeneity associated with Zeeman dephasing. This compact high-stability magnetic regulation scheme provides an engineering solution for DLCZ cold-atom quantum memories and offers design guidance for SERF magnetometers and space-borne ultracold atomic clock platforms.

## 2. Theory.

In experiments preparing entangled photons on a cold atom platform, the atoms are prepared with equal probability in the $|5S_{1/2}, F=1, m=\pm1,0\rangle$ three Zeeman sublevels as shown in Fig. 1, during the write process, a pulsed optical field blue-detuned by 20 MHz from the $|5S_{1/2}, F=1\rangle \rightarrow |5P_{1/2}, F'=2\rangle$ transition and with $\sigma^+$ polarization interacts with the atomic ensemble, generating $S_{-1,-1}$, $S_{-1,1}$, $S_{0,0}$, $S_{0,2}$, and $S_{1,1}$, five spin waves stored in the atomic ensemble. During readout, a pulsed optical field resonant with the $|5S_{1/2}, F=2\rangle \rightarrow |5P_{1/2}, F'=1\rangle$ transition and with $\sigma^-$ polarization retrieves $S_{-1,1}$, $S_{0,0}$, $S_{0,2}$, and $S_{1,1}$. The spin wave $S_{-1,-1}$ is not retrieved and therefore remains a dark state.

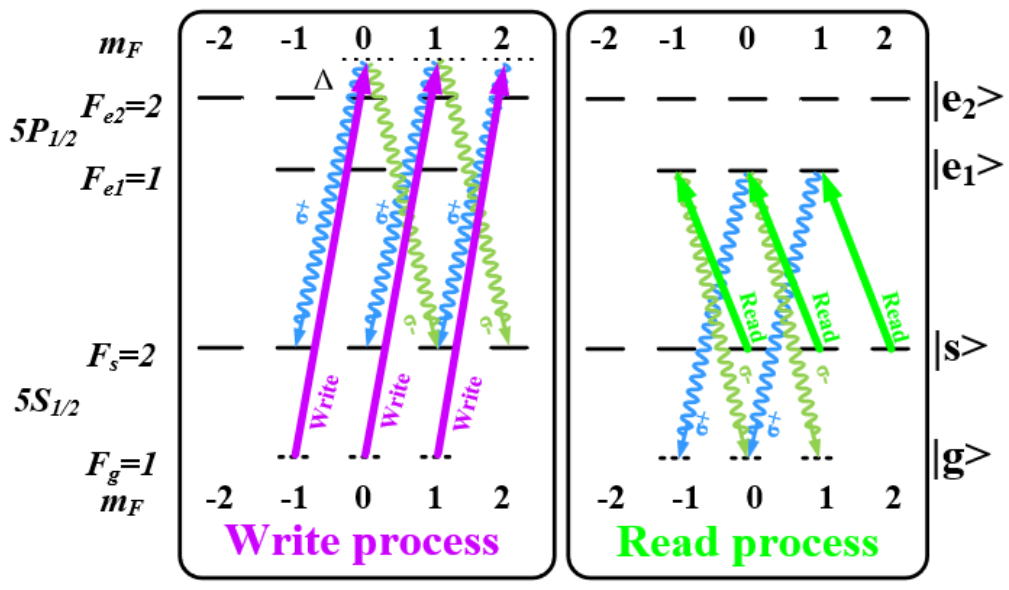

Fig. 1. Schematic energy-level structure of the write and read processes

The temporal evolution of an arbitrary spin-wave mode can be expressed as:

$$S_{m,m'}(z,t) = S_{m,m'}(z,0)\frac{\sum_{m=-1}^{0} p_{m,\alpha=1}^{\beta=\alpha}\sqrt{N_m}e^{-\frac{t^2}{2\tau_{m,m'}{}^2}}e^{-i\omega_{m,m'}t}e^{-\frac{t}{2\tau_D}}+\sum_{m=0}^{1} p_{m,\alpha=1}^{\beta=-\alpha}\sqrt{N_m}e^{-\frac{t^2}{2\tau_{m,m'}{}^2}}e^{-i\omega_{m,m'}t}e^{-\frac{t}{2\tau_D}}}{\sqrt{\left(\sum_{m=-1}^{0}\left|p_{m,\alpha=1}^{\beta=\alpha}\right|^2+\sum_{m=0}^{1}\left|p_{m,\alpha=1}^{\beta=-\alpha}\right|^2\right)}} \quad (1)$$

Here $N_m$ denotes the atomic population residing in each Zeeman sublevel $|5S_{1/2}, F=1, m=\pm1,0\rangle$. The atoms are uniformly distributed over the three sublevels with equal population for each sublevel. The quantities $m$ and $m'$ are the magnetic quantum numbers, while $p_{m_g,\alpha}^{\beta=\alpha} = \frac{C_{m_g\alpha(m_g+\alpha)}^{F^s 1F^e}}{C_{m_g\beta(m_g+\alpha)}^{F^g 1F^e}}$ and $p_{m_g,\alpha}^{\beta=-\alpha} = \frac{C_{(m_g-\alpha+\beta)\alpha(m_g+\beta)}^{F^s 1F^e}}{C_{m_g\beta(m_g+\beta)}^{F^g 1F^e}}$ denote the relative weights of the different spin-wave coherence terms. $C_{m_s\beta(m_s+\beta)}^{F^s 1F^e}$ $(C_{m_g\alpha(m_g+\alpha)}^{F^g 1F^e})$ is the Clebsch-Gordan coefficient for the $|F=s\rangle \rightarrow |F'=e_1\rangle$ $(|F=g\rangle \rightarrow |F'=e_2\rangle)$ transition. $\omega_{m,m'} = \frac{B_z\mu_B}{h}\left[g_g(m+m') - \delta g m\right]$ is the Larmor frequency of spin wave $S_{m,m'}(z,t)$, $\tau_{m,m'} = \frac{h}{(|g_g(m+m')-\delta g m'|\mu_B l B')}$ is the lifetime of spin wave $S_{m,m'}(z,t)$, and $\tau_D$ is the maximum storage time determined by atomic motion. The Landé factors of levels $|g\rangle$ and $|s\rangle$ are $g_g = 0.4998$ and $g_s = -0.5018$, thus $\delta g = g_{g+}g_s = -0.0020$. l denotes the length of atomic ensemble. In the theoretical model, the z direction is chosen along the quantization axis. The parameter $B'$ denotes the effective spatial gradient of the magnetic-field component along this selected axis.

The readout efficiency can be expressed as $\gamma(t) = \left|\langle S_{m,m'}(z,0)|S'_{m,m'}(z,t)\rangle\right|^2$. The retrieval efficiency can then be written as:

$$\gamma(t) = \gamma_0 \left| p_{-1,1} e^{-\frac{t^2}{2\tau_{-1,1}{}^2} - i\omega_{-1,1}t} + p_{1,1} e^{-\frac{t^2}{2\tau_{1,1}{}^2} - i\omega_{1,1}t} + p_{0,0} e^{-\frac{t^2}{2\tau_{0,0}{}^2} - i\omega_{0,0}t} + p_{0,2} e^{-\frac{t^2}{2\tau_{0,2}{}^2} - i\omega_{0,2}t} \right|^2 e^{-\frac{t}{\tau_D}} \quad (2)$$

where $\gamma_0$ is the readout efficiency at $t = 0$, and $p_{-1,1} = \frac{1}{10}$, $p_{1,1} = \frac{4}{10}$, $p_{0,0} = \frac{3}{10}$, and $p_{0,2} = \frac{2}{10}$. For $B_z = -2.5$ μT, the calculated curve in Fig. 2(a) decays with periodic oscillations, whereas for $B_z = 0$, the calculated curve in Fig. 2(b) exhibits a monotonic non-oscillatory decay.

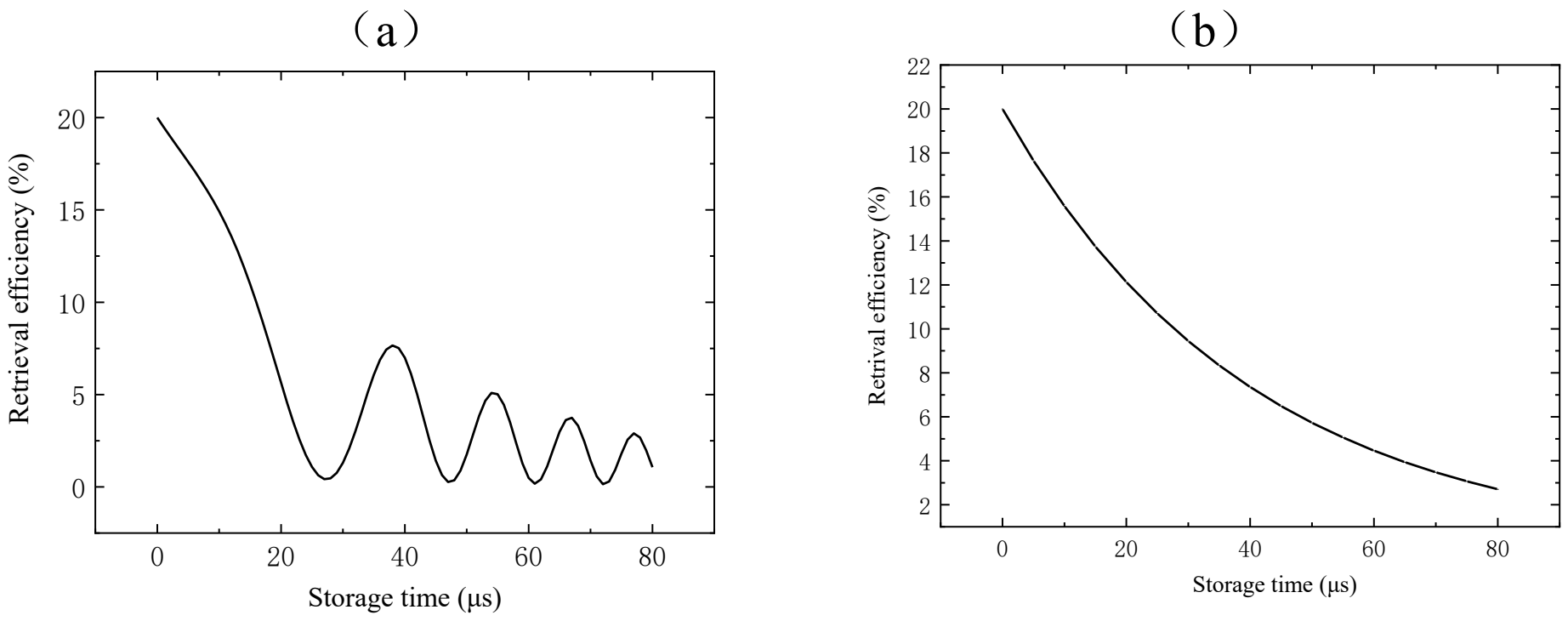

Fig. 2. Illustrative theoretical curves. (a) represents the variation process of the readout efficiency over time when there is no compensation magnetic field; (b) represents the variation process of the readout efficiency over time after the application of the compensation magnetic field.

3. Experimental Setup and Feedback Compensation.

The theoretical derivation above quantitatively proves that residual axial magnetic fields and spatial gradients induce differential Larmor frequencies, distort spin-wave interference and degrade retrieval efficiency. To suppress Zeeman decoherence and realize the ideal near-zero magnetic condition required by the model, we construct a closed-loop triaxial active magnetic compensation platform, whose overall architecture is illustrated in Fig. 3. The system consists of four core functional modules: orthogonal square Helmholtz coil assembly, multi-channel programmable DC power supply (HY-PMC), CH-3600 three-axis digital gaussmeter, and upper computer with integrated control algorithm. The power source outputs three independent adjustable currents to drive the coil group and generate three orthogonal compensation magnetic fields counteracting ambient geomagnetic interference. The Hall probe of the gaussmeter is rigidly fixed at the geometric center of the coil skeleton, enabling real-time synchronous sampling of three orthogonal field components *Bx, By, Bz* within the atomic working volume. Measured field data are continuously transmitted to the host computer via serial communication for closed-loop calculation. On the software side, a combined feedforward decoupling and incremental PI control strategy is executed iteratively. The controller compares real-time measured residual fields against the zero-field setpoint, computes required correction currents via the pre-calibrated inverse sensitivity matrix, and dynamically updates the power supply outputs to offset time-varying geomagnetic drifts. Through this real-time feedback loop, the magnetic field at the geometric-centre feedback location can be maintained in a stable near-zero regime; finite-volume spatial uniformity surrounding this point is evaluated independently in Sec. 4.2.

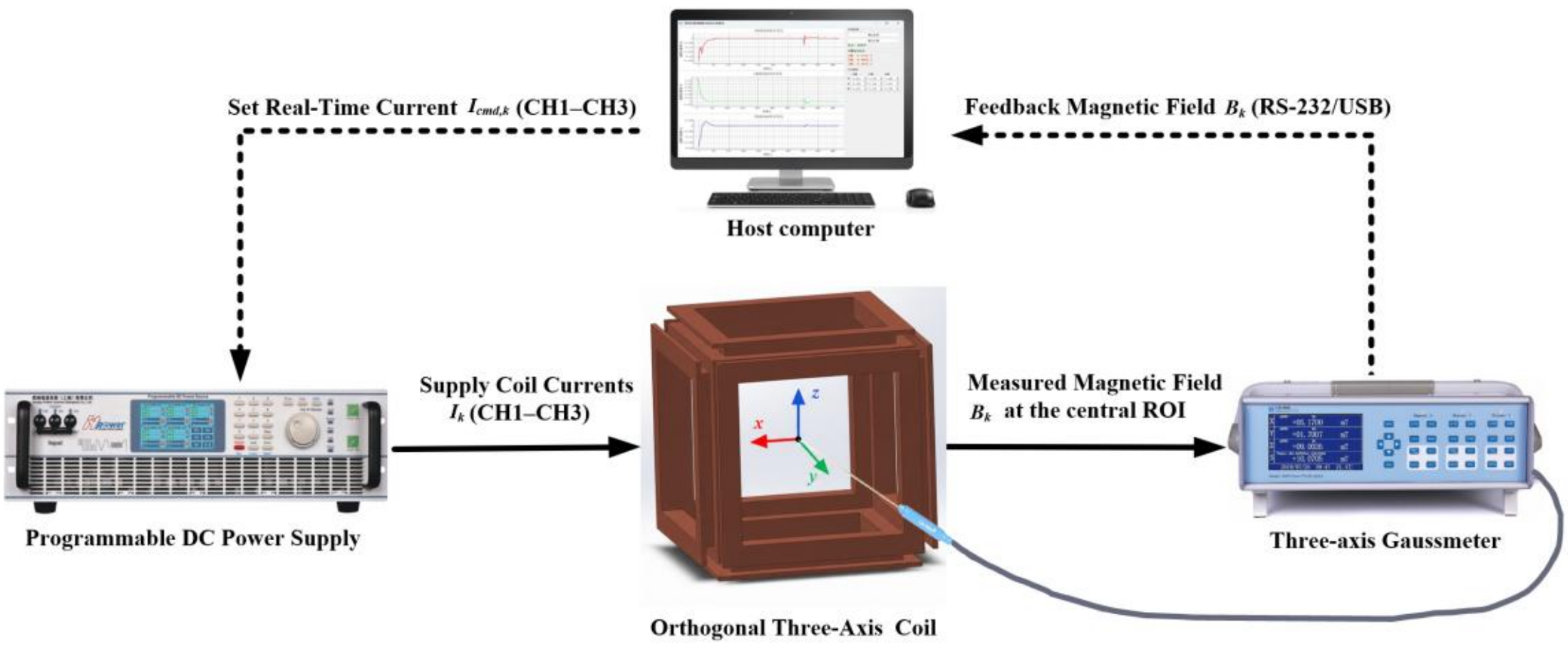

Fig. 3. Schematic diagram of the triaxial active magnetic compensation closed-loop measurement and control system.

This work adopts an integrally 3D-printed square Helmholtz coil framework with an effective side length of $a = 158\,\text{mm}$. The ambient background magnetic field mainly originates from geomagnetic signals and indoor ferromagnetic building components. We define the triaxial driving current vector $\boldsymbol{I} = [I_1; I_2; I_3]^T$ and measured residual field vector $\boldsymbol{B} = [B_x; B_y; B_z]^T$. The linear input-output mapping between driving currents and detected magnetic fields follows[39]:

$$\boldsymbol{B} = \boldsymbol{M}\boldsymbol{I} + \boldsymbol{B_0} \tag{3}$$

where the $3 \times 3$ matrix $\mathbf{M}$ is the coil-field sensitivity matrix, and constant vector $\mathbf{B}_0$ denotes static background offset including geomagnetism and structural residual magnetism. At system startup, the initial uncompensated field was measured as $\mathrm{B_x} = -94.1\ \mu\text{T}$, $\mathrm{B_y} = 128.3\ \mu\text{T}$, and $\mathrm{B_z} = -190.7\ \mu\text{T}$, corresponding to a total background magnitude $\|B_0\| \approx 248.4\ \mu\text{T}$. For closed-loop compensation, we focus on the incremental relation between current adjustments and field variations, written as:

$$\Delta\boldsymbol{B} = \boldsymbol{M} \cdot \Delta\boldsymbol{I} \tag{4}$$

Since the DC power supplies only output unipolar currents, a 2.0 A fixed bias is superimposed on all three channels to realize bidirectional field offset. The full sensitivity matrix $\boldsymbol{M}$ is calibrated via multi-point current sweeping and least-squares fitting, as expressed in Eq. (5):

$$\boldsymbol{M} = \begin{bmatrix} \boldsymbol{M_{x1}} & \boldsymbol{M_{x2}} & \boldsymbol{M_{x3}} \\ \boldsymbol{M_{y1}} & \boldsymbol{M_{y2}} & \boldsymbol{M_{y3}} \\ \boldsymbol{M_{z1}} & \boldsymbol{M_{z2}} & \boldsymbol{M_{z3}} \end{bmatrix} = \begin{bmatrix} 3.49 & 88.77 & -60.84 \\ 0.36 & 59.54 & 93.66 \\ -108.57 & 3.32 & -0.86 \end{bmatrix} \left(\frac{\mu T}{A}\right) \tag{5}$$

To eliminate inherent cross-axis coupling contained in $\boldsymbol{M}$, a feedforward decoupling scheme based on the inverse sensitivity matrix $\boldsymbol{M}^{-1}$ is implemented. All PI control increments are first calculated in magnetic field error space and then converted to physical current corrections via matrix inversion, as shown in Eq. (6):

$$\Delta\boldsymbol{I} = \boldsymbol{M}^{-1} \cdot \Delta\boldsymbol{u_B} \tag{6}$$

where $\Delta\boldsymbol{u_B} = \left[\Delta u_{Bx}, \Delta u_{By}, \Delta u_{Bz}\right]^T$ represents the virtual field correction vector at each discrete update step $k$, denoted as $\Delta\boldsymbol{u_{B,k}}$ and $\Delta I_k$ respectively. The measured condition number was $\kappa_2(\boldsymbol{M}) = 1.047$, so the inverse mapping is numerically well conditioned. Each column of $\boldsymbol{M}$ represents the three-axis magnetic-field vector generated per unit current in one physical channel. The simultaneous nonzero responses along multiple sensor axes reflect the combined effects of coil orientation, assembly deviations, and sensor-coordinate alignment. These calibrated vector responses are retained in the inverse mapping rather than approximated by independent scalar channel gains.

The three calibrated actuation-vector magnitudes were 108.63, 106.94 and 111.69 μT $\text{A}^{-1}$, while the singular values of $\boldsymbol{M}$ ranged only from 106.81 to 111.86 μT $\text{A}^{-1}$. Together with the pairwise angles near 90°, these values show that the integrated coil assembly provides a nearly orthogonal and gain-balanced three-dimensional actuation basis rather than merely an invertible numerical matrix.

For near-zero magnetic control, the reference target is set as $B_{ref} = 0$. At the $k$-th eligible sampling instant, the field error vector and its infinity-norm evaluation criterion are defined as:

$$e_k = B_{ref} - B_k, s_k = \| e_k \|_\infty \quad (7)$$

The infinity norm $s_k$ extracts the maximum single-axis field deviation, such that all three components must satisfy the threshold condition simultaneously to trigger state switching. Under the ACTIVE compensation state, incremental PI control operates in virtual field space according to Eq. (8):

$$\Delta u_{B,k} = K_p(e_k - e_{k-1}) + K_i e_k \quad (8)$$

Fixed discrete gains $K_p = 0.4$ and $K_i = 0.1$ are adopted, with a minimum eligible update interval of 0.40 s as the system timing baseline. We introduce a dual-threshold hysteresis state machine $q_k \in \{\text{ACTIVE, HOLD}\}$ to suppress frequent current toggling induced by the gaussmeter's 0.1 μT quantization step (1 LSB). The entry threshold for HOLD mode is set to $\delta_{enter} = 0.2$ μT (2 LSBs), and the exit threshold is $\delta_{exit} = 0.4$ μT (4 LSBs) to form a stable hysteresis band. The complete state transition rule is formalized as Eq. (9):

$$q_k = \begin{cases} \text{HOLD} & q_{k-1} = \text{ACTIVE} \wedge s_k \le \delta_{enter} \\ \text{ACTIVE} & q_{k-1} = \text{HOLD} \wedge s_k \ge \delta_{exit} \\ q_{k-1} & \text{otherwise} \end{cases} \quad (9)$$

Under ACTIVE mode, virtual field increments are converted to current adjustments via $\mathbf{M}^{-1}$ and clipped within the 0~4A actuator range, following Eq. (10)

$$\Delta I_k = \mathbf{M}^{-1} \Delta u_{B,k},$$
$$I_{prop,k} = \text{clip}(I_{cmd,k-1} + \Delta I_k) \quad (10)$$

If the clipped candidate current $I_{prop,k}$ is identical to the previously issued command $I_{cmd,k-1}$, serial transmission to the power supply is skipped. Otherwise, three-channel currents are updated synchronously in an atomic transaction. New current setpoints, operating modes and error logs are only stored after successful communication feedback; transmission failures discard all pending updates and roll back to the last valid controller state to eliminate field mismatch caused by incomplete current configuration.

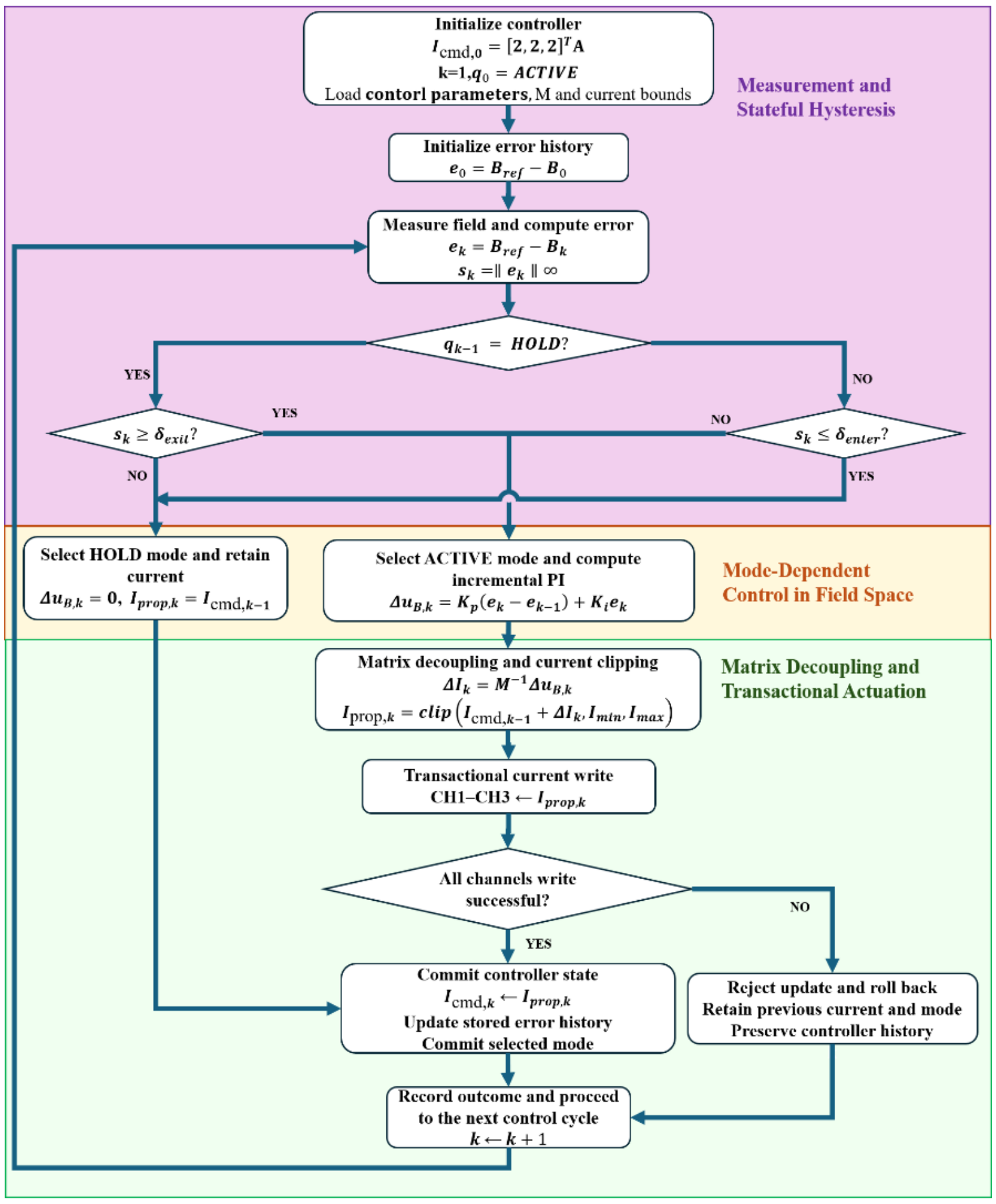

Fig.4. Flowchart of matrix-decoupled incremental PI control with dual persistent hysteresis thresholds.

The full closed-loop workflow integrating magnetic sampling, error judgment, hysteresis state switching, decoupled incremental PI regulation, current limiting and transactional output is summarized in Fig. 4. The dual-hysteresis logic suppresses unnecessary actuator switching within the ultraweak steady-field region. All PI corrections are decoupled via $\boldsymbol{M}^{-1}$ and clamped to the hardware current limit before transmission, while the HOLD state freezes output currents to avoid repetitive minor adjustments. Every control cycle starts with three-axis magnetic sampling and infinity-norm error calculation; the 0.2/0.4 μT thresholds determine whether the system remains in static HOLD or executes iterative PI compensation. All updated controller states are committed only upon successful serial write feedback or confirmed no-write judgment.

4. Results and Discussion.

4.1 Long-Time Temporal Stabilization Performance

We first evaluate the repeatability and dynamic compensation characteristics of the proposed triaxial decoupled control system via three independent long-duration experiments, each running over 2 h under identical ambient geomagnetic interference. A threshold criterion is defined uniformly for all tests: the system is regarded as reaching a stable target magnetic regime only when the total field norm $\| \mathbf{B} \|_2$ remains below the specified value for a continuous 30 s. Across the three runs, the initial vector-field magnitude was (216.7 ± 33.5) μT. The system reached the sustained 2 μT band after (34.3 ± 2.3) s and the sustained 1 μT band after (130.6 ± 28.0) s. For the unified post-stabilization window of 90–120 min, the time-weighted mean residual field is 0.879 ± 0.232 μT with an RMS value of $1.004 \pm 0.254$ μT. All three groups successfully trigger the HOLD state, with the first entry time ranging from 157.5 s to 564.4 s, and no channel current saturation occurs throughout all measurements, which verifies the excellent repeatable convergence performance of the integrated compensation framework down to the microtesla level. We select the trial with an initial background field magnitude of 248.4 μT for staged dynamic analysis (Fig. 5). This dataset contains three clearly separated working phases and features the longest uninterrupted HOLD episode (~714.4 s) among all tests, serving as a typical representative to illustrate the full closed-loop regulation process rather than a universal average of all experimental batches. At the initial sampling instant of this run, the raw triaxial field components are $B_x = -94.1$ μT, $B_y = 128.3$ μT, and $B_z = -190.7$ μT, which jointly produce the strong initial geomagnetic offset of 248.4 μT.

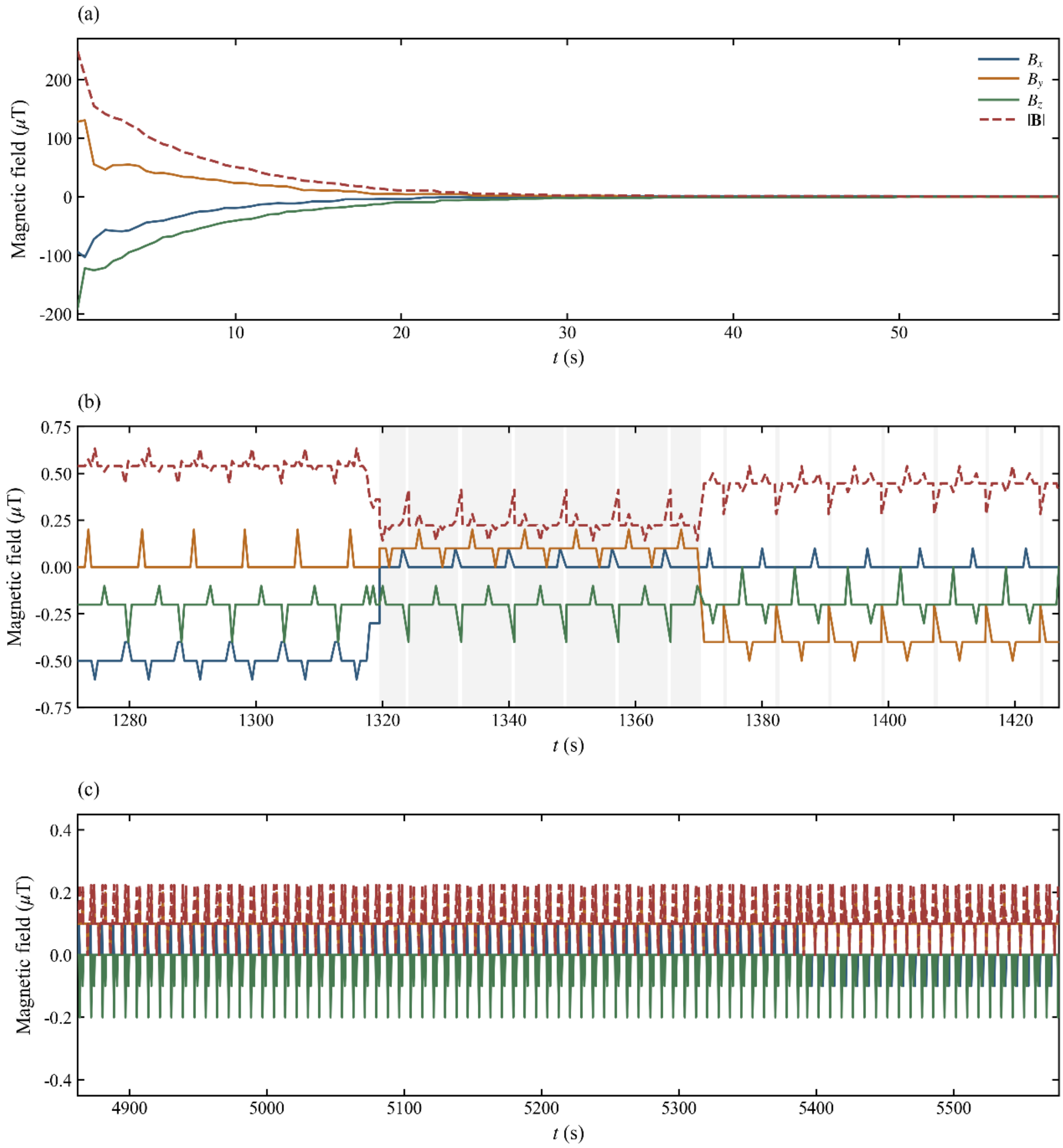

Fig. 5. Dynamic convergence and anti-interference responses of the three-axis active magnetic compensation system. (a) Stage I: Rapid compensation of initial geomagnetic field; (b) Stage II: Intermittent fine regulation with alternating ACTIVE/HOLD states; (c) Stage III: Continuous steady-state HOLD period.

Phase I covers the time window of 0.484–59.625 s and corresponds to fast transient suppression. Large field deviations generate substantial incremental PI correction outputs, which rapidly counteract the strong initial geomagnetic field and pull residual signals down to the microtesla range. The decay speed slows progressively as the residual field approaches zero, because the control increment shrinks synchronously with field error magnitude. Phase II spans 1271.8–1426.9 s and characterizes the medium-term fine-tuning stage with frequent mode switching. Within this window, the mean and peak values of $\| \mathbf{B} \|_2$ are 0.4 μT and 0.6 μT, and the HOLD state accounts for 37.1% of the whole duration. Phase III extracts a 714.4 s continuous HOLD segment (4862.2–5576.5 s), during which the controller locks all output currents without any update commands. The average residual field over this ultra-long steady period is merely 0.112μT, with a maximum value limited to 0.224μT. The three-stage curve distinctly distinguishes the system's three core capabilities: rapid large-offset cancellation, intermittent precise trimming, and ultra-long near-zero field retention. Notably, the dual-threshold state machine adopts the infinity-norm error metric $s_k = \| \boldsymbol{e_k} \|_\infty$ instead of the total vector norm $\| \mathbf{B} \|_2$ to judge mode switching. The threshold condition is satisfied only when every single-axis deviation falls within the limit, which means the synthetic field magnitude may slightly exceed 0.2 μT even if the HOLD entry criterion is met. To quantitatively assess low-frequency drift in the 714.4 s steady window, power spectral density and overlapping Allan deviation analyses are conducted. For averaging times ranging from 13.9 s to 178.2 s, the Allan deviations of $B_x, B_y, B_z$ all stay below 5 nT, proving outstanding long-term stability of the compensated central magnetic zone.

4.2 Spatial Uniformity of the 3D Near-Zero Working Volume

Temporal measurements characterize field stability at the fixed feedback probe, whereas spatial uniformity requires a separate assessment of the field surrounding the centre. The spatial scan sampled 27 positions on a three-dimensional lattice with each coordinate taking values of -5, 0, and +5 mm under fixed logged current commands. Five global-centre measurements were used to form a component-wise, piecewise-linear temporal reference, with the nearest endpoint retained outside the reference interval. For the n-th lattice measurement acquired at position r_n and time t_n, the signed centre-referenced field difference is defined as

$$\Delta \mathbf{B}_n^{\mathrm{obs}} = \mathbf{B}_{\mathrm{meas}}(\mathbf{r}_n, t_n) - \widehat{\mathbf{B}}_{\mathrm{c}}(t_n), \qquad n = 1, \ldots, 27 \tag{11}$$

where the first term is the measured three-component magnetic-field vector and the second term is the estimated global-centre field at the same acquisition time. Upper-plane centre revisits were retained as consistency checks rather than independently reset to zero. The correction therefore preserves the signs of all three components and the estimated differences between measurement planes, subject to the assumption that the temporal drift is approximately spatially common over the scanned region.

The corrected vectors were reconstructed jointly using a centre-constrained quadratic vector model. To improve numerical conditioning, the spatial coordinate is normalized by L = 5 mm. For each field component, the reconstruction and the nine-term polynomial basis are written as

$$\begin{gathered} \widehat{\Delta B}_i(\mathbf{r}) = \mathbf{c}_i^{\mathrm{T}} \boldsymbol{\Phi}(\boldsymbol{\xi}), \qquad i \in \{x, y, z\}, \qquad \boldsymbol{\xi} = \mathbf{r}/L, \quad L = 5 \text{ mm} \\ \boldsymbol{\Phi}(\boldsymbol{\xi}) = \left[\xi_x, \xi_y, \xi_z, \xi_x^2, \xi_y^2, \xi_z^2, \xi_x\xi_y, \xi_x\xi_z, \xi_y\xi_z\right]^{\mathrm{T}} \end{gathered} \tag{12}$$

Here $\mathbf{c}_i$ is the coefficient vector of the i-th reconstructed component. Omission of the constant term enforces a zero centre-referenced field difference at r = 0 by construction. This is only a reference condition and should not be interpreted as evidence that the absolute magnetic field is exactly zero at the geometric centre.

Because the scan volume contains no intentional current source and is treated as a region of approximately uniform permeability, the source-free magnetostatic relations provide a useful physical prior. They are introduced as a soft regularization rather than as exact constraints. After stacking the three coefficient vectors into $\mathbf{c}$, the fitted coefficients are obtained from

$$\hat{\mathbf{c}} = \arg\min_{\mathbf{c}} \left[ \| \mathbf{D}\mathbf{c} - \mathbf{b} \|_2^2 + \frac{\lambda_{\mathrm{M}}}{N_{\mathrm{P}}} \| \mathbf{P}_{\mathrm{M}}\mathbf{c} \|_2^2 + \lambda_{\mathrm{R}} \mathbf{c}^{\mathrm{T}} \mathbf{W} \mathbf{c} \right] \tag{13}$$

where $\boldsymbol{c} \in \mathbb{R}^{27}$ stacks the three nine-element coefficient vectors, $\boldsymbol{b} \in \mathbb{R}^{81}$ contains the 81 centre-referenced scalar components measured at the 27 lattice positions, and $\boldsymbol{D} \in \mathbb{R}^{81\times 27}$ is the corresponding polynomial design matrix. The matrix $\boldsymbol{P}_{\mathrm{M}} \in \mathbb{R}^{500\times 27}$ evaluates one divergence and three curl components at $N_c = 125$ collocation positions arranged on a $5 \times 5 \times 5$ grid in the normalized coordinate domain, so that $N_P = 4N_c = 500$. The derivatives in $\boldsymbol{P}_{\mathrm{M}}$ are evaluated with respect to the normalized coordinates $\boldsymbol{\xi} = \boldsymbol{r}/L$; because the same spatial scale $L = 5$mm is applied to all three axes, the zero-divergence and zero-curl conditions retain their directional form. The matrix $\boldsymbol{W} \in \mathbb{R}^{27\times 27}$ is diagonal, with weights of 1 for linear coefficients and 4 for quadratic coefficients. The regularization parameters $\lambda_{\mathrm{M}} = 1.077217345$ and $\lambda_{\mathrm{R}} = 1.583819609$ were selected by leave-one-site-out cross-validation to minimize the predictive vector-field RMS error. The Maxwell term therefore acts only as a soft physical regularization that suppresses strongly nonphysical spatial variation while retaining the structure supported by the measurements; the resulting field is neither an exact interpolant nor an exactly Maxwell-constrained solution.

The scalar quantity used for spatial evaluation is the magnitude of the reconstructed field difference relative to the centre, rather than the absolute residual-field magnitude. Two geometrical regions of interest (ROIs) are considered: a spherical region of radius 5 mm and an elongated prolate ellipsoid representing a cigar-shaped application volume. They are defined by

$$\begin{gathered} d(\mathbf{r}) = \| \widehat{\Delta\mathbf{B}}(\mathbf{r}) \|_2 \\ \Omega_{\mathrm{s}} = \{\mathbf{r}\colon x^2 + y^2 + z^2 \le (5\ \mathrm{mm})^2\} \\ \Omega_{\mathrm{e}} = \left\{\mathbf{r}\colon \frac{x^2}{(5\ \mathrm{mm})^2} + \frac{y^2}{(2.5\ \mathrm{mm})^2} + \frac{z^2}{(2.5\ \mathrm{mm})^2} \le 1\right\} \end{gathered} \tag{14}$$

The ellipsoid has full axis lengths of 10, 5, and 5 mm, with its major axis along x. It is an illustrative elongated application ROI and is not a directly measured atomic-cloud boundary. For either region $\Omega$, spatial variation is quantified by the RMS centre-referenced field difference and by the fraction satisfying the 5 μT criterion:

$$\begin{gathered} B_{\mathrm{RMS},\Omega} = \left[\frac{1}{V_\Omega}\int_\Omega d^2(\mathbf{r})\,\mathrm{d}V\right]^{1/2} \\ \eta_{5,\Omega} = \frac{100\%}{V_\Omega}\int_\Omega \mathbb{1}\,[d(\mathbf{r}) \le 5\ \mu\mathrm{T}]\,\mathrm{d}V \end{gathered} \tag{15}$$

where $V_\Omega$ is the geometrical volume of the selected ROI and $\mathbb{1}$ denotes the indicator function. Numerical evaluation gives an RMS variation of 5.65 μT and a qualifying volume fraction of 67.5% for the spherical ROI. For the elongated ellipsoidal ROI, the corresponding values are 2.68 μT and 93.7%. The higher qualifying fraction in the elongated ROI results primarily from its smaller transverse extent and its alignment along the less sensitive x direction. It therefore describes the dependence of the same reconstructed field on the selected evaluation geometry rather than an improvement in the underlying compensation performance. The 5 μT value is introduced here only as a common spatial-variation criterion for comparing the reconstructed finite-volume field distribution. It is distinct from the sub-1 μT temporal residual-field target at the central feedback position and from the 0.2/0.4 μT hysteresis thresholds used for controller state switching.

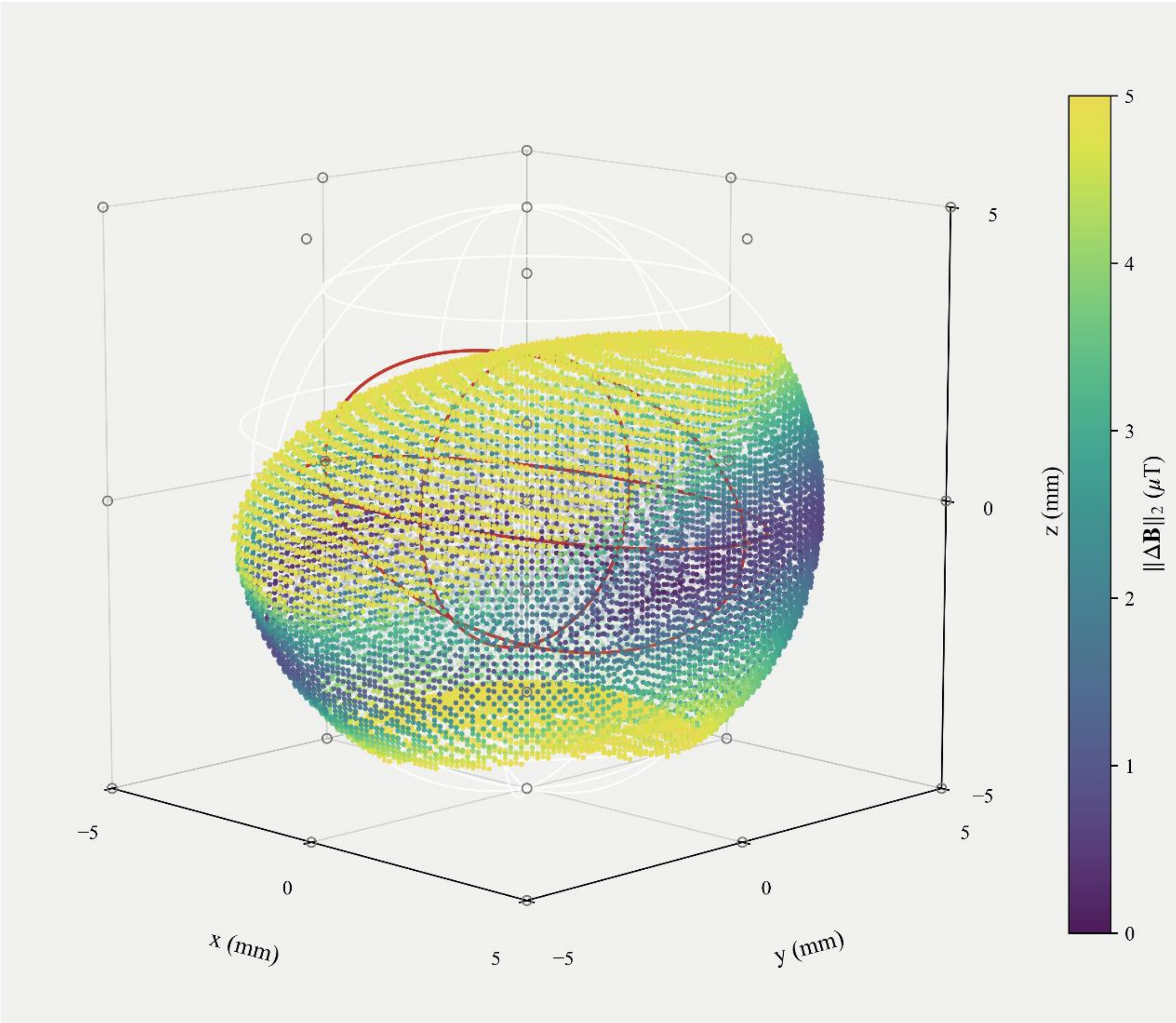

Fig. 6. Three-dimensional model estimate of the region satisfying d(r) <= 5 μT inside the spherical ROI of radius 5 mm. Coloured points are subsampled reconstruction-grid positions rather than individual measurements, and their colours indicate the centre-referenced field-difference magnitude on a common 0-5 μT scale. The white framework outlines the spherical ROI, the red curves show principal sections of the prolate ellipsoidal ROI with semi-axes (5, 2.5, 2.5) mm, and the open grey markers indicate the 27 measured lattice positions. The ROI curves are geometrical references rather than magnetic-field isosurfaces.

Fig. 6 shows that the reconstructed lower-variation volume is distinctly anisotropic. The qualified region occupies a substantial part of the central volume but is compressed towards positive z, whereas it extends farther towards negative z. This asymmetry is consistent with the stronger vertical spatial variation identified by the reconstructed vector field. The spherical ROI therefore includes a considerable peripheral volume in which the centre-referenced variation exceeds 5 μT, producing the 67.5% qualifying fraction. By contrast, the elongated ROI excludes much of the strongly varying transverse periphery while retaining the full 10 mm extent along x, reducing the RMS variation to 2.68 μT and increasing the model-estimated qualifying fraction to 93.7%. These values are model-based volume statistics and should not be interpreted as experimentally verified uniform-volume coverage.

To further identify the directional origin of this three-dimensional anisotropy, the same reconstructed field is examined on orthogonal central and offset planes in Fig. 7.

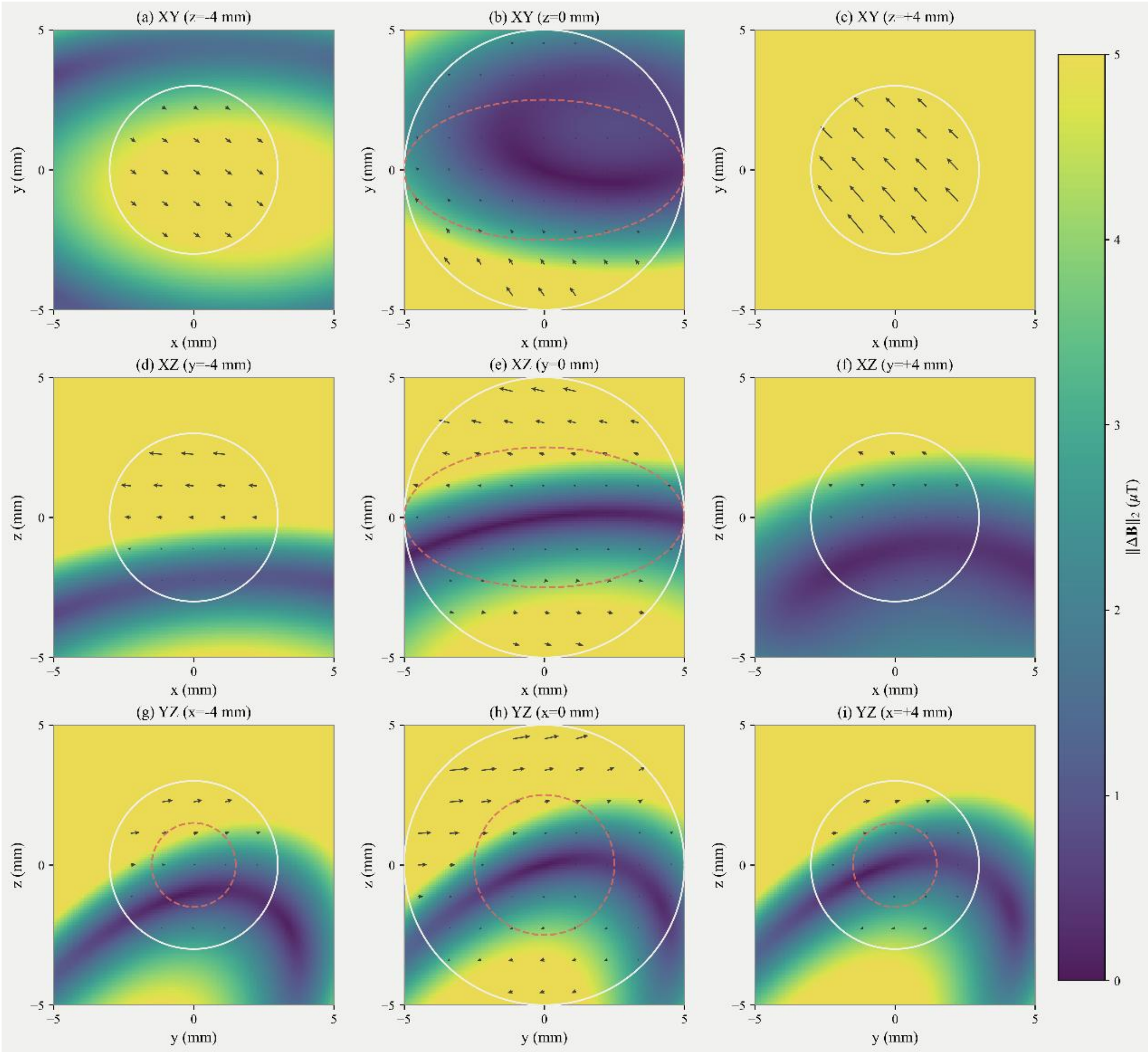

Fig. 7. Planar magnitude and vector-projection maps from the same regularized quadratic reconstruction as Fig. 6. Rows show XY, XZ, and YZ planes, while the three columns correspond to offsets of -4, 0, and +4 mm along the respective normal axes. Colours encode d(r) on the same 0-5 µT scale used in Fig. 6; values above 5 µT are saturated at the upper colour limit. Arrows show the in-plane projections of the reconstructed centre-referenced vector field. White solid curves denote intersections with the spherical ROI and red dashed curves denote intersections with the ellipsoidal ROI; ellipsoidal sections are absent when the selected plane does not intersect the ellipsoid. Continuous maps and arrows are model evaluations.

The slice maps in Fig. 7 further reveal the directional origin of the reconstructed anisotropy. The central XY plane contains a broader lower-variation region than the central XZ and YZ planes, indicating that the strongest spatial change occurs along the vertical direction. In particular, the displayed XY plane at z = +4 mm remains above 5 µT throughout the shown area, whereas the z = -4 mm plane still contains appreciable lower-valued regions. The background colour represents the full three-component magnitude, while the arrows contain only the projection in the displayed plane; arrow length therefore need not follow the background colour when the out-of-plane component is appreciable. The ±4 mm sections are reconstructed evaluation planes between the measured -5, 0, and +5 mm lattice planes and are not additional spatial measurements.

For direct comparison of the three central planes, Eq. (15) is also evaluated over circular disks of radius 5 mm by replacing the volume integrals with area integrals. On a uniform numerical grid, the central XY disk has an RMS variation of 2.75 μT, a grid-evaluated maximum of 8.60 μT, and a qualifying area fraction of 89.8%. The corresponding values are 5.59 μT, 14.98 μT, and 63.3% for the XZ disk, and 6.29 μT, 17.71 μT, and 62.0% for the YZ disk. These planar area fractions and the three-dimensional volume fractions above have different statistical domains and should not be used interchangeably.

To quantify the directional variation independently of the finite ROI size, the local centre Jacobian and the norm of its response along each spatial coordinate are defined as

$$(\mathbf{J}_0)_{ij} = \left.\frac{\partial \widehat{\Delta B}_i}{\partial r_j}\right|_{\mathbf{r}=\mathbf{0}}, \qquad g_j^{\mathrm{loc}} = \| \mathbf{J}_0 \mathbf{e}_j \|_2, \qquad j \in \{x, y, z\} \tag{16}$$

where $\mathbf{e}_j$ is the unit vector along the j-th spatial axis and derivatives are taken with respect to the physical coordinates in millimetres. The reconstructed local variation rates are $g_x$ = 0.269, $g_y$ = 0.634, and $g_z$ = 2.098 μT mm$^{-1}$. Thus, the z-direction rate is approximately 7.81 times the x-direction rate and 3.31 times the y-direction rate. These quantities combine the spatial changes of all three reconstructed magnetic-field components and therefore represent vector-field inhomogeneity rather than single-component axial gradients. In particular, $g_z^{\mathrm{loc}}$ should not be directly substituted for the effective gradient $B'$ appearing in the spin-wave model of Sec. 2. A direct comparison with $B'$ would require projection of the reconstructed Jacobian onto the experimentally defined quantization axis.

Finite-distance variation from the centre to the spherical boundary is characterized separately by the directional secant rate

$$g_{j,\pm}^{\mathrm{sec}}(R) = \frac{\left\| \widehat{\Delta \mathbf{B}}(\pm R\mathbf{e}_j) - \widehat{\Delta \mathbf{B}}(\mathbf{0}) \right\|_2}{R}, \qquad R = 5\ \mathrm{mm} \tag{17}$$

At R = 5 mm, the reconstructed secant rates are 0.078 and 0.563 μT mm$^{-1}$ along +x and -x, 0.504 and 1.700 μT mm$^{-1}$ along +y and -y, and 2.969 and 1.252 μT mm$^{-1}$ along +z and -z, respectively. The unequal positive- and negative-direction values arise from the quadratic terms of the reconstruction and quantify the observed asymmetry without uniquely identifying its physical origin.

The spatial reconstruction should therefore be interpreted as a conditional estimate of the measured field distribution. The vector training RMSE over all 27 lattice positions is approximately 8.34 μT, and fitted and measured field-difference magnitudes fall on opposite sides of the 5 μT threshold at six positions, all outside the spherical ROI. Residual temporal drift, positioning and probe-orientation errors, and model approximation may contribute to the remaining discrepancy. Consequently, the reported volume and area fractions quantify the fitted reconstruction rather than independently verified experimental coverage. The centre-point temporal stabilization demonstrated in Sec. 4.1 must therefore be distinguished from finite-volume spatial uniformity and from a direct experimental demonstration of improved cold-atom coherence.

## 5. Conclusion.

This paper proposes a compact triaxial active magnetic compensation system equipped with spatial feedforward inverse matrix decoupling and dual-threshold hysteretic incremental PI regulation, aiming to suppress Zeeman-induced decoherence in cold atom quantum memories. Mechanically, an integrated 3D printed high-orthogonal square Helmholtz skeleton is fabricated to cut down inherent inter-axis magnetic flux leakage at the structural layer. Algorithmically, a least-squares-calibrated inverse sensitivity matrix realizes feedforward pre-compensation of coil assembly cross-coupling, and a hysteresis deadband control module is embedded into the PI loop to eliminate frequent current adjustments triggered by digital gaussmeter quantization noise near zero field, forming a complete spatiotemporal dual magnetic interference suppression architecture.

Systematic long-duration closed-loop experiments verify the system's dynamic and steady-state performance. Starting from an initial composite geomagnetic background of 216.7 ± 33.5 μT, the system achieves stable sub-1 μT residual field within 130.6 ± 28.0 s on average. During a representative 714.4 s uninterrupted HOLD stage, the mean residual field is 0.112 μT with a maximum value of 0.224 μT, and the overlapping Allan deviations of all three axes remain below 5 nT over averaging times of 13.9–178.2 s, demonstrating low-frequency long-term stability at the central feedback location. Spatial field mapping based on a drift-referenced, centre-constrained quadratic vector reconstruction with soft Maxwell regularization further reveals pronounced anisotropy around the centre. Within the 5 mm-radius spherical ROI, the reconstructed RMS centre-referenced variation is 5.65 μT and 67.5% of the model volume satisfies the 5 μT criterion. For the prolate ellipsoidal ROI with full axis lengths of 10, 5, and 5 mm, the corresponding RMS variation is 2.68 μT and the qualifying fraction is 93.7%. The reconstructed z-direction variation is the strongest and therefore constrains the usable three-dimensional low-variation volume. The remaining peripheral inhomogeneity indicates that the present compact coil geometry still limits the extent of this volume. These spatial statistics are conditional model estimates rather than independently verified experimental coverage, but together with the temporal results they demonstrate the feasibility of localized magnetic regulation for the central atomic working region. Although the compact coil layout suffers weak anti-interference ability against external magnetic gradients at peripheral boundaries, the proposed decoupling control strategy can provide effective localized magnetic regulation for the central atomic cloud, validating the practicability of miniaturized near-zero magnetic platforms. The spatiotemporal joint suppression mechanism simultaneously mitigates static Zeeman splitting and Larmor inhomogeneous broadening, fundamentally reducing magnetic-induced spin-wave phase noise.

This compact, high-stability magnetic regulation scheme delivers a full set of hardware realization methods and supporting theoretical foundations for DLCZ cold atom quantum memories, and also provides systematic design references for SERF atomic magnetometers and space-borne ultracold atomic clocks. Future work will focus on optimizing axial coil geometric structures to reduce vertical magnetic gradients, alongside adaptive PI tuning algorithms, to further expand the homogeneous near-zero working volume and strengthen the system's ability to resist time-varying ambient magnetic disturbances.

## References

[1] Hammerer K, Sørensen A S and Polzik E S 2010 *Reviews of Modern Physics* **82** 1041–1093

[2] Pu Y-F, Zhang S, Wu Y-K, Jiang N, Chang W, Li C and Duan L-M 2021 *Nature Photonics* **15** 374–378

[3] Hétet G, Hosseini M, Sparkes B M, Oblak D, Lam P K and Buchler B C 2008 *Optics Letters* **33**

[4] Marangos M F A I J P 2005 *REVIEWS OF MODERN PHYSICS* **77**

[5] Wang M, Wang S, Xu Z, Ma T, Li Y, Li S and Wang H 2022 *Applied Physics B* **128**

[6] Wang S, Xu Z, Wang D, Wen Y, Wang M, Zhou P, Yuan L, Li S and Wang H 2019 *Optics Express* **27** 27409–27419

[7] Li Y, Wang W, Wu Q, Chen Y, Sun C, Wang H and Qiao W J a e-p 2023
[8] Yang S-J, Wang X-J, Bao X-H and Pan J-W 2016 *Nature Photonics* **10** 381–384
[9] Sun C, Li Y, Hou Y b, Wang M j, Li S j and Wang H 2023 *Advanced Quantum Technologies* **6**
[10] Hänni J, Rodríguez-Moldes A E, Appas F, Wengerowsky S, Lago-Rivera D, Teller M, Grandi S and de Riedmatten H 2025 *Physical Review X* **15**
[11] Heller L, Farrera P, Heinze G and de Riedmatten H 2020 *Physical Review Letters* **124** 1–6
[12] Duan Z-X, Wu W-T, Lin Y-T and Yang S-J 2022 *Review of Scientific Instruments* **93**
[13] Dudin Y O, Jenkins S D, Zhao R, Matsukevich D N, Kuzmich A and Kennedy T A 2009 *Phys Rev Lett* **103** 1–4
[14] Zhu 朱 S-L 诗 2026 *Chinese Physics Letters* **43**
[15] Kouassi N, Doumbia V, Boka K, Tuo Z, Grodji O D F, Kassamba A A and Zillé A F 2021 *Journal of Earth System Science* **130**
[16] Hammer M D, Cox G A, Brown W J, Beggan C D and Finlay C C 2021 *Earth, Planets and Space* **73**
[17] Xu X, Wu W, Wang L, Liu W and Han Q 2026 *Journal of Materials Chemistry C*
[18] Wang K, Ma D, Li S, Gao Y and Sun J 2023 *Sensors and Actuators A: Physical* **349**
[19] Xu X and Liu Y 2024 *Review of Scientific Instruments* **95**
[20] Li J, Quan W, Zhou B, Wang Z, Lu J, Hu Z, Liu G and Fang J 2018 *IEEE Sensors Journal* **18** 8198–8207
[21] Zhao Z, Xu X, Liu W, Wu W and Han Q 2026 *Composites Communications* **61**
[22] Yang Z q, Li Y b, Liu X k, Li Y p, Cui P l, Zheng S q and Zhou X x 2025 *IEEE Transactions on Automation Science and Engineering* **22** 1006–1015
[23] Xu X, Han C, Liu Y and Zhang X 2026 *IEEE Transactions on Instrumentation and Measurement* **75** 1–9
[24] Liu Z, Zhang Q, Pan M, Shan Q, Geng Y, Guan F, Chen D and Tian W 2017 *IEEE Geoscience and Remote Sensing Letters* **14** 48–51
[25] Wang S, Lu J, Lu F, Lu C and Zhang X 2022 *IEEE Transactions on Instrumentation and Measurement* **71** 1–9
[26] T. Brys S C, M. Daum, P. Fierlinger, D. George, R. Henneck, Z. Hochman, M. Kasprzak, K. Kohlik, K. Kirch, M. Kuzniak, G. Kuehne, A. Pichlmaier, A. Siodmok, A. Szelc, and L. Tanner 2005 *Journal of Research of the National Institute of Standards and Technology*
[27] Raven H P C v 2022 A new Caesium quantum gas microscope with precise magnetic field control Ludwig-Maximilians-Universität München)
[28] Fu Y, Hao X, Jia R, Yin W, Yu X, Feng X and Zhang D 2025 *IEEE Sensors Journal* **25** 19040–19051
[29] Vaez M, Sahrai M and Vafafard A 2025 *Journal of Physics B: Atomic, Molecular and Optical Physics* **58**
[30] Sun J, Gan Z, Shi Y, Zhou W, Wang K and Cheng J 2025 *IEEE Transactions on Industrial Electronics* **72** 14900–14908
[31] ; Y X, ; Z L, ; Q Z, Guan; F, ; Z Y, Huang; B, ; M P, ; J H, ; Z C, ; Q D, ; X Q and Tang Y 2023 *Review of Scientific Instruments*
[32] Qin L, Hang C, Li W and Zhao X 2026 *Chinese Physics Letters* **43**
[33] Holmes N, Tierney T M, Leggett J, Boto E, Mellor S, Roberts G, Hill R M, Shah V, Barnes G R, Brookes M J and Bowtell R 2019 *Scientific Reports* **9**

[34] Tian K, Zhang X, Shi M, Yang J, Cao F, Wang F, Shi Z, Wang K, Zheng S and Liu G 2025 *IEEE Transactions on Industrial Electronics* **72** 1065–1075

[35] Han B, Yang J, Zhang X, Shi M, Yuan S and Wang L 2023 *IEEE Transactions on Industrial Electronics* **70** 2057–2065

[36] Hu X, Ge Z, Liu S, Li G, Huang Y and Wang Y 2017 *Plos One* **12**

[37] Batista D S, Granziera F, Tosin M C and de Melo L F 2018 *IEEE Transactions on Aerospace and Electronic Systems* **54** 392–403

[38] Hobson P J, Vovrosh J, Stray B, Packer M, Winch J, Holmes N, Hayati F, McGovern K, Bowtell R, Brookes M J, Bongs K, Fromhold T M and Holynski M 2022 *Scientific Reports* **12**

[39] Ramdas V and Sethunadh R 2022 *IEEE Access* **10** 81846–81855